\documentclass[twocolumn,prb]{revtex4}
\usepackage{epsfig,amssymb,amsmath}
\begin{document}
\title{Structure of the breakpoint region in CVC of the intrinsic Josephson junctions}
\author{ Yu.M.Shukrinov~$^{1,2}$}
\author{F.Mahfouzi~$^{1}$ }
\author{M.Suzuki~$^{2,3}$ }
\address{$^{1}$ BLTP, JINR, Dubna, Moscow Region, 141980, Russia\\
$^{2}$Photonics and Electronics Science and Engineering Center, Kyoto University, Kyoto 615-8510, Japan\\
$^{3}$Department of Electronic Science and Engineering, Kyoto University, Kyoto 615-8510, Japan}
\date{\today}
\begin{abstract}
A fine structure of the breakpoint region in the current-voltage characteristics of the coupled intrinsic
Josephson junctions in the layered superconductors is found.  We establish a correspondence between the features
in the current-voltage characteristics and the character of the charge oscillations in superconducting layers in
the stack and explain the origin of the breakpoint region structure.
\end{abstract}
\maketitle A system of  Josephson junctions attracts a great interest from both scientific and practical sides.
To describe it,  a system of coupled nonlinear equations is used, which might be solved numerically only.
Recently observed\cite{ozyuzer}  powerful enough coherent radiation from the stack of the intrinsic Josephson
junctions (IJJ) in  layered superconductor BSCCO opens new perspectives for different  applications.

In Refs.\cite{sust1,prl,prb} we studied the multiple branch structure of the current-voltage characteristics
(CVC) of IJJ and showed that the branches have the breakpoint (BP) and a breakpoint region (BPR) before
transition to the another branch. The BP current is determined by the creation of the longitudinal plasma waves
(LPW) with a definite wave number $k$, which depends on the coupling parameter $\alpha$, the dissipation
parameter $\beta$, the number of junctions in the stack, and the boundary conditions.  We generalized the
McCumber-Stewards dependence of the return current for the case of IJJ in the HTSC and the oscillation of the
breakpoint current as a function of parameters $\alpha$ and  $\beta$  was found. Based on the idea of the
parametric resonance in the stack of IJJ, we explained this oscillation as a result of the creation of
longitudinal plasma waves with different wave numbers. A good qualitative agreement of the results of modeling
of the $\alpha\beta$-dependence of the BP current with the results of simulation was obtained.  We demonstrated
that the $\alpha\beta$-dependence of the BP current is an instrument to determine the mode of LPW created at the
breakpoint in the stacks with a different number of junctions. So, the breakpoint in the CVC was explained, but
the nature and features of the breakpoint region  were not clear at that time.

In this paper we study theoretically the phase dynamics of a stack of  IJJ in the high-$T_c$ superconductors.
The CVC of IJJ are numerically calculated in the framework of capacitively coupled Josephson junctions model
with diffusion current.\cite{machida00,physC2} From the results of the numerical simulations we predict a fine
structure of the breakpoint region in the CVC.  A correspondence between the features in the CVC and the
character of the charge oscillations on different superconducting layers is established and  the origin of the
breakpoint region structure is explained. In the BPR the plasma mode is a stable solution of the system and this
fact might be used in some applications, particularly,  in high frequency devices such as THz oscillators and
mixers. We consider that the breakpoint phenomenon might be used to understand the mechanism of the powerful
coherent radiation from the stack of the IJJ in BSCCO.

To find the CVC of the stack of IJJ, we solve a system of dynamical equations for the gauge-invariant phase
differences $\varphi_l(\tau)= \theta_{l+1}(\tau)-\theta_{l}(\tau)-\frac{2e}{\hbar}\int^{l+1}_{l}dz
A_{z}(z,\tau)$ between superconducting layers ($S$-layers) for the stacks with a different number of IJJ in the
framework of the capacitively coupled Josephson junctions model with diffusion current (CCJJ+DC
model)\cite{machida00,physC2}, where $\theta_{l}$ is the phase of the order parameter in the S-layer $l$, $A_z$
is the vector potential in the barrier. We use a dimensionless time $\tau = t\omega_p$, where $\omega_{p}$ is
the plasma frequency $\omega_{p}=\sqrt{2eI_c/\hbar C}$, ${I_c}$ is the critical current and $C$ is the
capacitance. The system of equations has a form $\partial^2\varphi_l/\partial \tau^2 =
\sum\sb{l'}\,A_{ll'}[I-\sin\varphi_{l'}-\beta\partial \varphi_{l'}/\partial \tau]$ with matrix A given in
Ref.\cite{prb} for periodic and nonperiodic boundary conditions (BC). In our simulations we measure the voltage
in units of $V_0=\hbar\omega_p/(2e)$ and the current in units of the $I_c$. The CVC and time dependence of the
charge oscillation in the S-layers are simulated at $\alpha = 1$ and $\beta = 0.2$ and periodic BC. The details
concerning the model and numerical procedure are presented in Refs.\cite{physC2,prl,prb}
\begin{figure}[ht]
 \centering
\includegraphics[height=55mm]{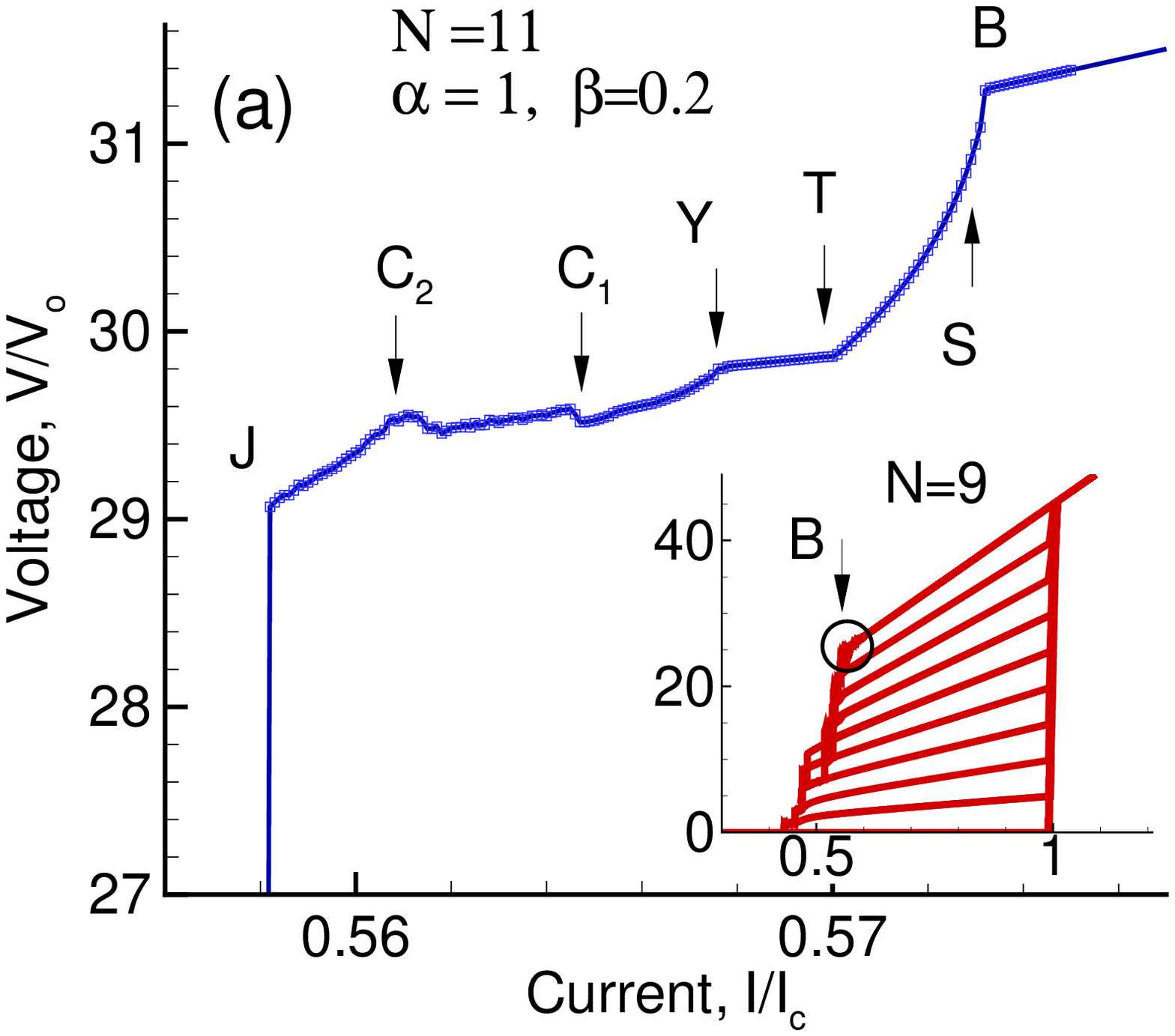}
\includegraphics[height=50mm]{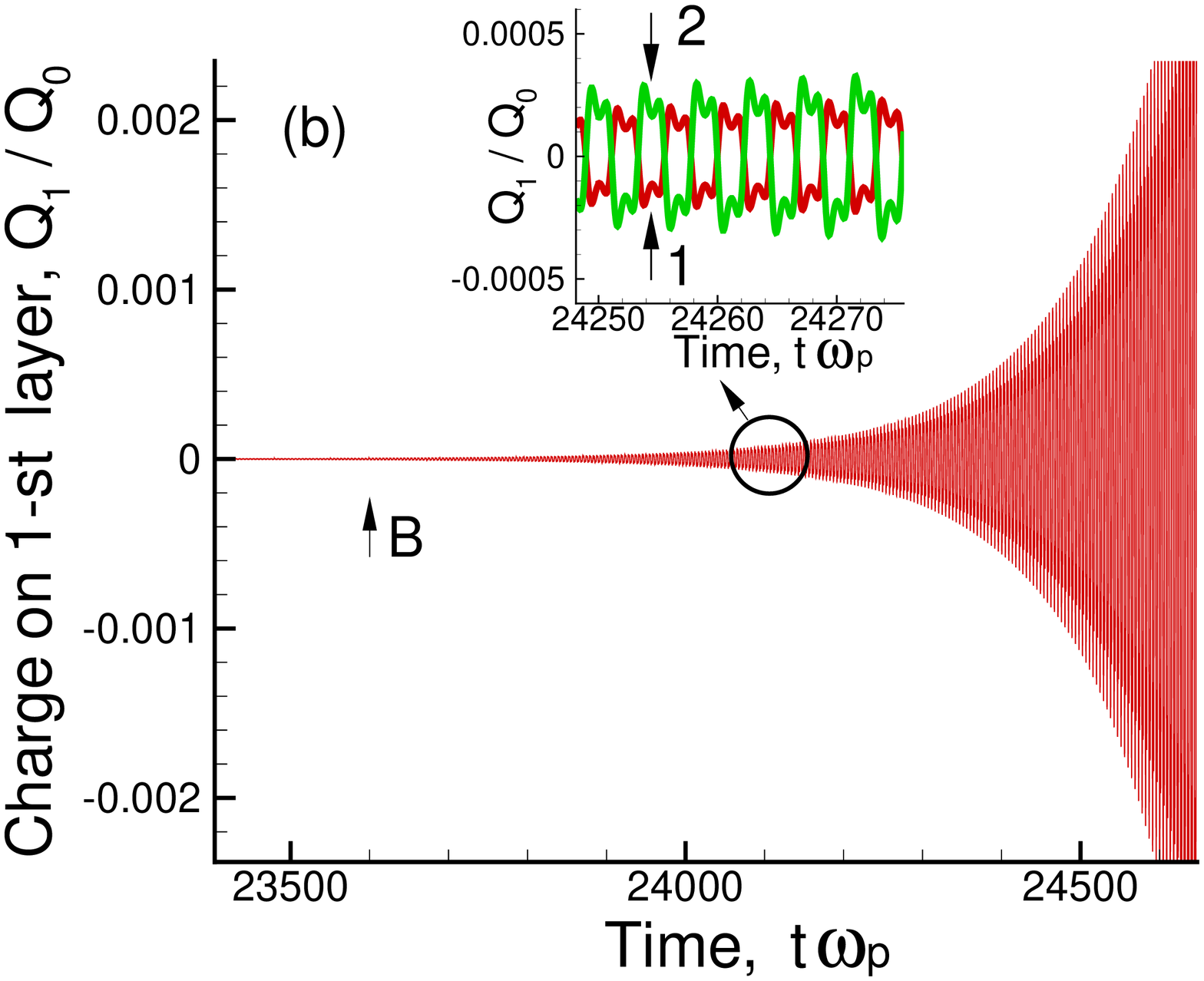}
\caption{(Color online)(a) The  fine structure of the BPR in the outermost branch in the CVC of the stack with
11 IJJ. The insert shows a total branch structure in the CVC of the stack with 9 IJJ and  the BP location. (b)
Charge oscillation in the first layer in the beginning of B-S part of the BPR. The inset shows the oscillations
in the layers 1 and 2.}
 \label{1}
\end{figure}
In the inset to Fig.~\ref{1}a we show the result of simulation of the CVC for a stack with 9 IJJ. It
demonstrates an equidistant multiple branch structure in the hysteresis region and the breakpoint location. The
transition from the state where the outermost branch with all junctions is in a rotating state
(R-state)\cite{matsumoto99} to a state with some junctions in an oscillating state (O-state ) happens in a
resonance region, where the LPW with a definite wave number is created.  In the stack with 11 IJJ  with
mentioned above values of $\alpha=1$ and $\beta=0.2$, the LPW with $k = 10 \pi /11d$ is created, where $d$ is
the lattice period along the c-axis.\cite{prl,prb} In Fig.~\ref{1}a we can see that the CVC exhibits a fine
structure in the BPR. The BPR has different parts with a different dependence of the voltage on the bias
current. For a given stack we can see clear five parts in the CVC separated by some points  $\emph{$T, Y, C_1,
C_2$}$, where the character of the CVC is changed. The letters $\emph{B}$ and $\emph{J}$ show the breakpoint and
jumping point to the another branch. The points $\emph{S}$ on the CVC will be discussed later.

To find the origin of this structure  and explain its features, we study time dependence of the charge in the
superconducting layers.  Using Maxwell equation $\emph{div} (\varepsilon\varepsilon_0 E) = \rho$, where
$\varepsilon$ and $\varepsilon_0$ are relative dielectric and electric constants, we express the charge density
$Q_l$ (we call it just charge) in the S-layer $l$ by the voltages $V_{l}$ and $V_{l+1}$ in the neighbor
insulating layers $Q_l=Q_0 \alpha (V_{l+1}-V_{l})$, where $Q_0 = \varepsilon \varepsilon _0 V_0/r_D^2$,  and
$r_D$ is Debay screening length. Solution of the system of dynamical equations for the gauge-invariant phase
differences  between S-layers gives us the voltages $V_{l}$ in all junctions in the stack, and it allows us to
investigate the time dependence of the charge on each S-layer.

The  "time dependence" actually consists of  time and bias current variation. We solve the system of dynamical
equations for phase differences at fixed value of bias current $\emph{I}$ in some time interval $(0, T_m)$ of
dimensionless time $\tau = t\omega_p$ with the time step $\delta \tau$, where $t$ is a real time. This interval
is used for time averaging procedure. Then we change the bias current by $\delta I$, and repeat the same
procedure for the current $I+\delta I$ in new time interval $(T_m, 2T_m)$. The values of the phase and
derivative of phase in the end of the first time interval are used as initial conditions for second time
interval and so on. In our simulations we put $T_m=250$, $\delta \tau=0.05$, $\delta I=0.0001$ and total
recorded time was calculated as $\tau+T_m(I_0 -I)/\delta I$, where $I_0$ is an initial value of the bias current
for time dependence recording.
\begin{figure}[ht]
 \centering
\includegraphics[height=60mm]{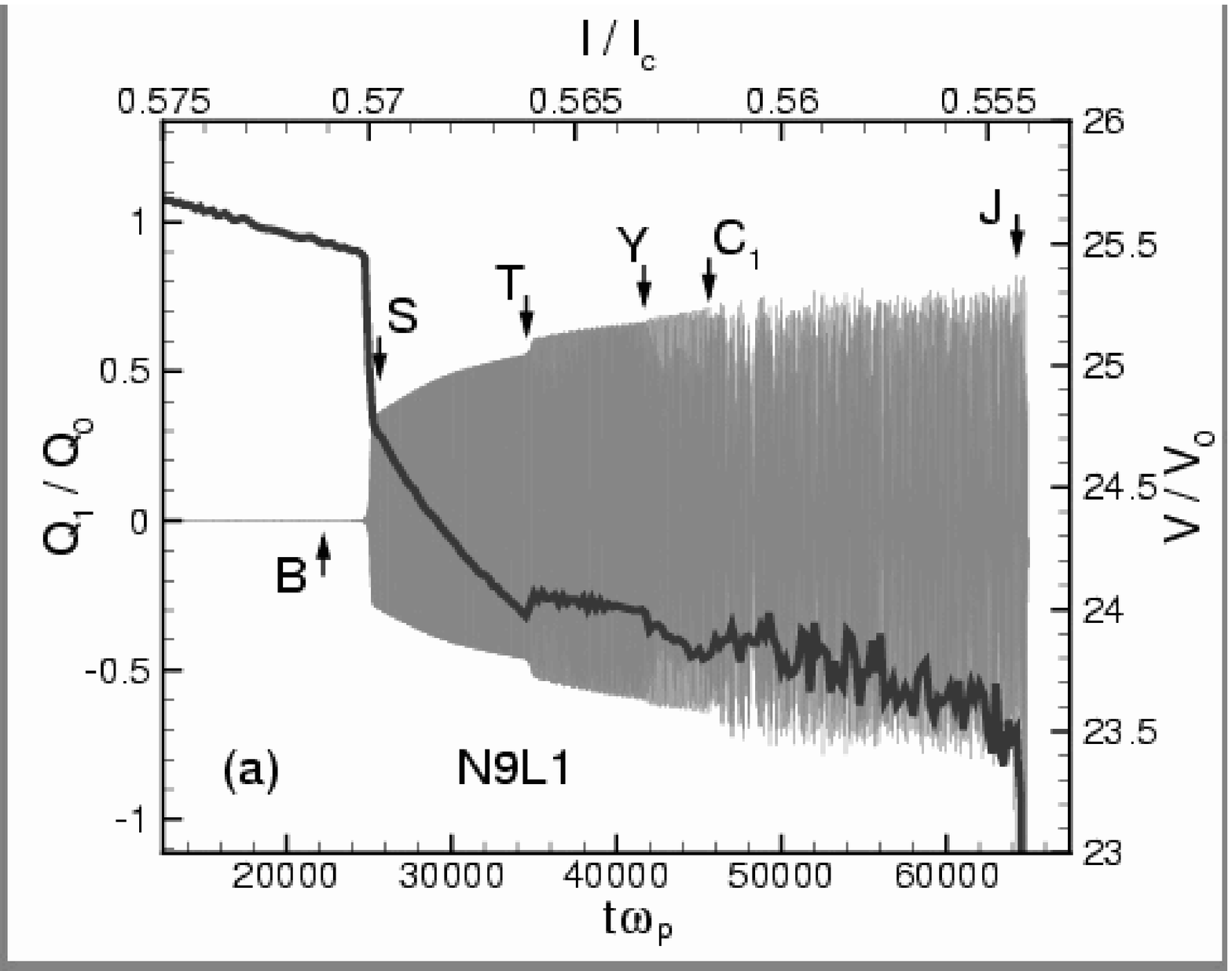}
\includegraphics[height=60mm]{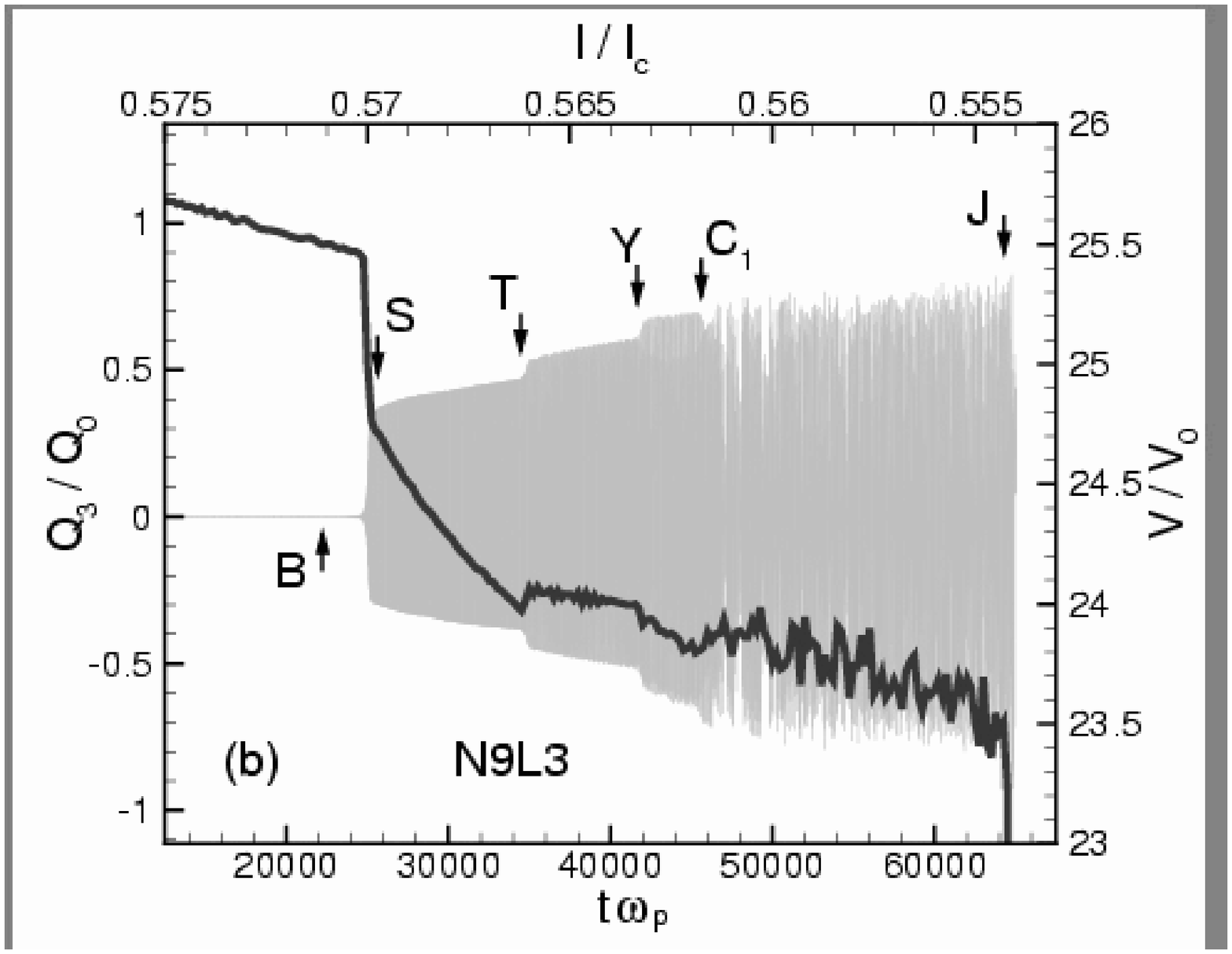}
\includegraphics[height=60mm]{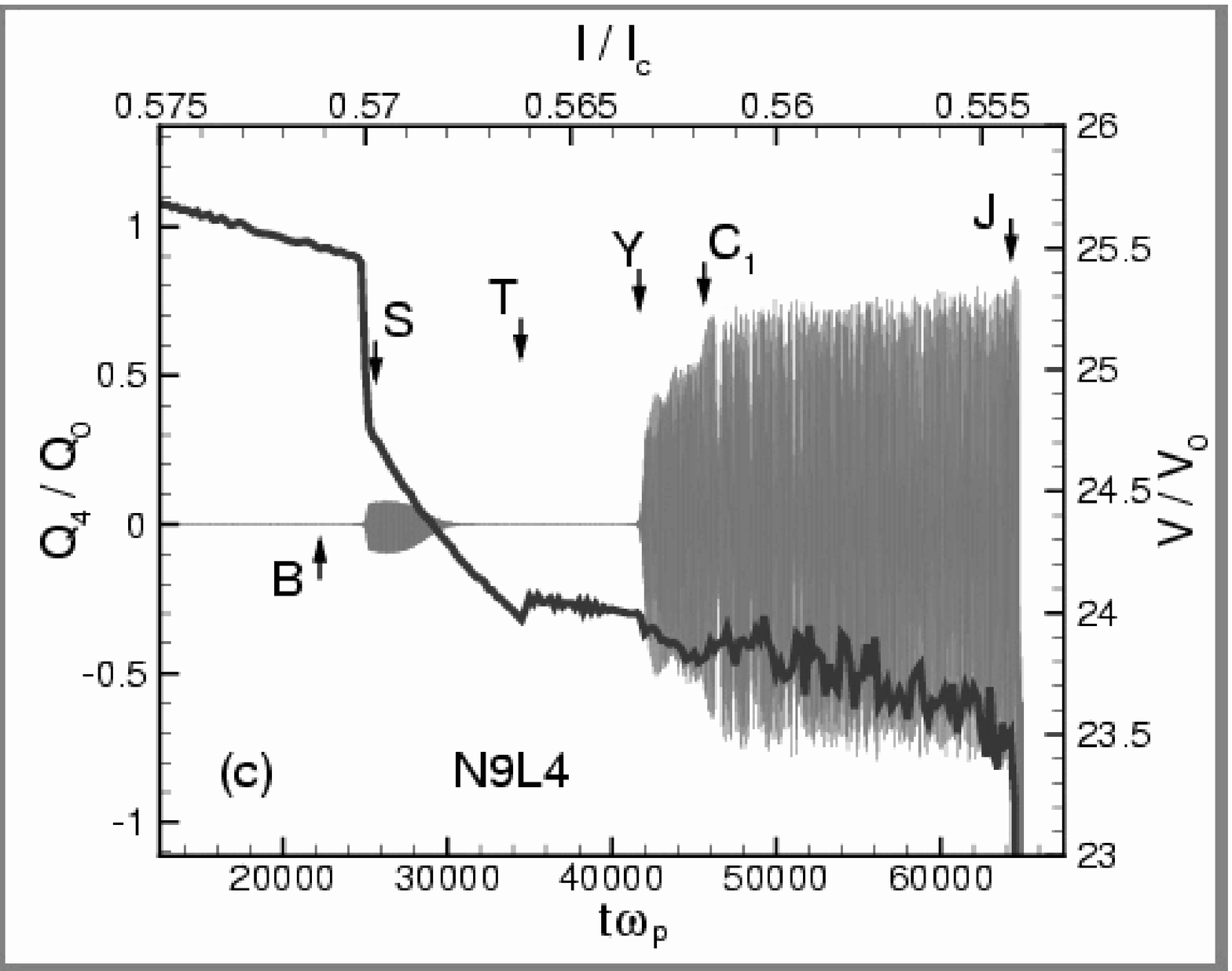}
\caption{(Color online) (a) Profile of time dependence of the charge in the S-layers in the stack with 9 IJJ:
(a) In the first layer. (b) In the third layer. (c) In the fourth layer. Red line shows the CVC of this stack in
the BPR.}
 \label{2}
\end{figure}

In the Fig.~\ref{2} the profile of time dependence of the  charge in the S-layers in the stack with 9 IJJ is
combined with the CVC of the outermost branch at $\alpha=1$, $\beta=0.2$ and periodic BC. We show oscillations
in the first (a), third (b) and fourth (c) layers only, because in the other layers its character close to the
behavior in these layers. To avoid large data files in the procedure of time dependence recording and to present
the CVC and time dependence with the same precision, we have limited the averaging time interval by $T_m=250$.
It leads to more smeared CVC in this figure, than the one shown in Fig.~\ref{1}a ($T_m=25000$), but the main
features of the CVC are preserved. Moreover, such an averaging also allows us to find one more characteristic
point $\emph{S}$ within the BPR.
\begin{figure}[ht]
 \centering
    \includegraphics[height=60mm]{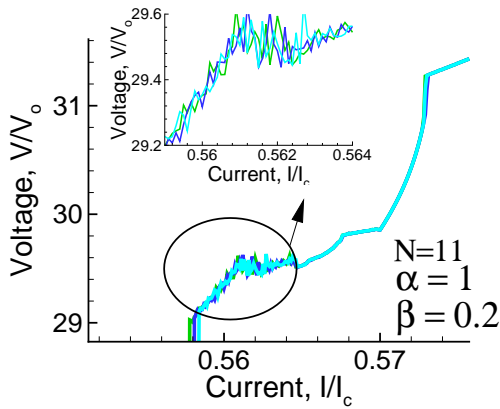}
\caption{(Color online) Superposition of the three CVC with the amplitudes of noise in the current $10^{-6}$,
$10^{-8}$ and $10^{-13}$. } \label{3}
\end{figure}

We have investigated the influence of the noise in bias current on the CVC, which is generated arbitrary in each
IJJs . The noise is produced  by random number generator and it's amplitude is normalized  to the critical
current value $I_c$. Fig.~\ref{3} presents the CVC of the outermost branch  in the vicinity of the breakpoint
for three values of noise amplitude: $10^{-6}$, $10^{-8}$ and $10^{-13}$ of a stack with 11 IJJ. We see that
first four parts of the BPR from the point $\emph{B}$ till the point $\emph{$C_1$}$ are not affected by noise in
current practically. But its strong effect has a place in the $\emph{$C_1-J$}$ region. The noise in bias current
smears the CVC essentially here because four first parts of the BPR demonstrate a behavior, close to the regular
one, but the part $\emph{$C_1-J$}$ shows the chaotic behavior. The voltage value is very sensitive  to the small
changes in the value of current here.

As we can see in the Fig.~\ref{2}, the features of the CVC are in correlation with the features of time
dependence of the charge on the S-layer. This fact can be explained by the idea of the parametric resonance at
point B.  As was shown in Ref.\cite{Koyama96}, the system of equations for CCJJ has a solution corresponding to
the LPW propagating along the c axis.  A frequency of the LPW at $I = 0$ and $\beta = 0$ is
$\omega_{LPW}(k)=\omega_p \sqrt{1+2\alpha(1-\cos kd)}$, where $k$ is wave vector of the LPW. At point B the
Josephson oscillations excite the LPW by their periodical actions. The frequency of Josephson oscillations is
determined by the voltage value in the junction, so at $\omega_J=2 \omega_{LPW}$ the parametric resonance is
realized and the LPW is created.

The sharp increase of the oscillation amplitude in the first layer of the stack with 9 IJJ in the beginning of
the B-S range is demonstrated in Fig.~\ref{1}b, where the LPW  with wave vector $k=8\pi/9d$ is created. This
mode  is close to the $\pi$-mode ($k=\pi /d$). In the $\pi$-mode charges on the nearest lattice points oscillate
in the opposite directions and the inset shows by numbers 1 and 2 that the charges oscillate  in the layers 1
and 2 in opposite directions too. But the amplitude of oscillations in the layer 2 is smaller than in the layer
1, which demonstrate the difference of this mode from the $\pi$- mode.
\begin{figure}[ht]
 \centering
 \includegraphics[height=60mm]{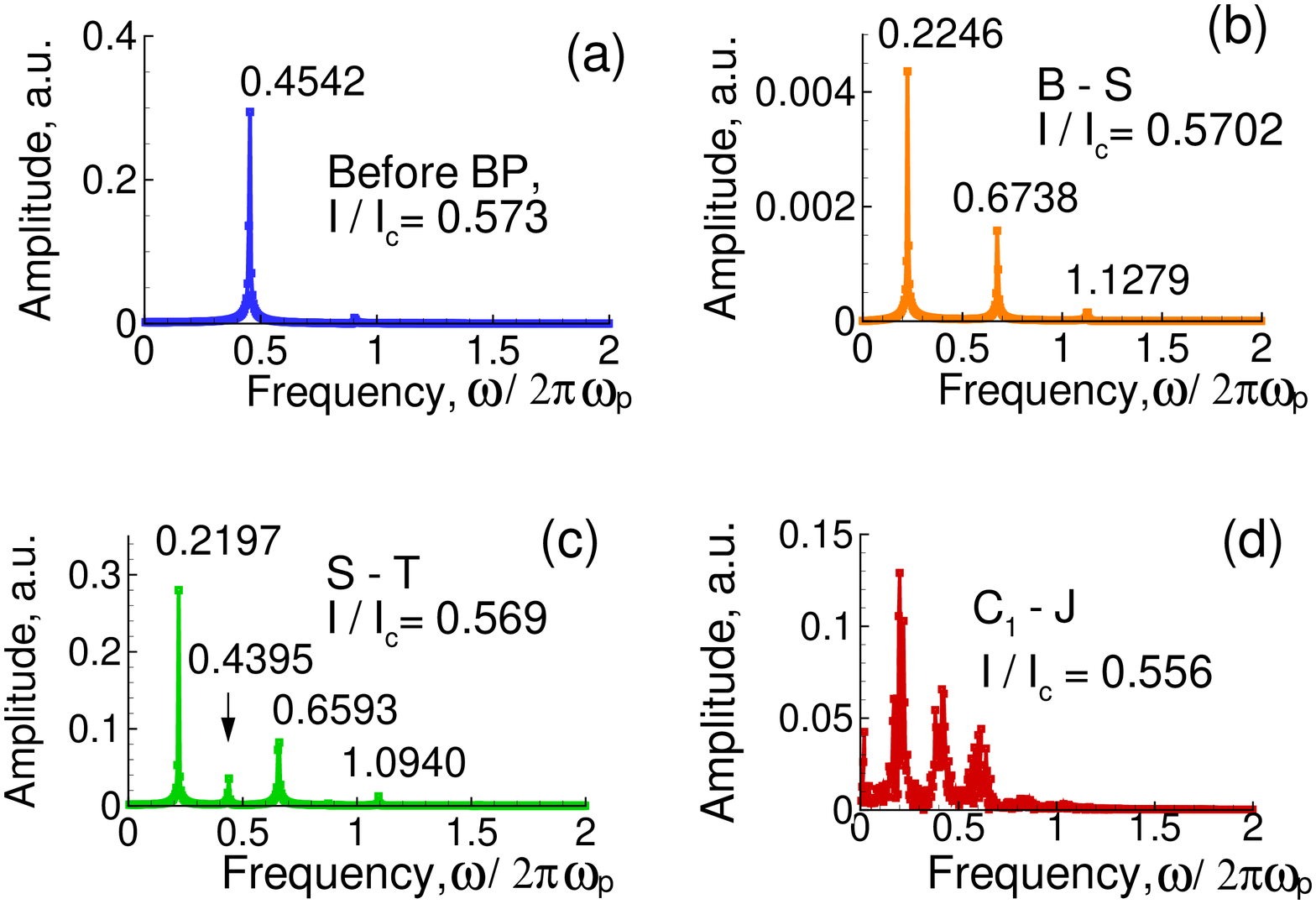}
\caption{(Color online) Results of FFT analysis of time dependence of the charge in the third S-layer in the
different parts of BPR}
 \label{4}
\end{figure}
\begin{figure}[ht]
 \centering
 \includegraphics[height=60mm]{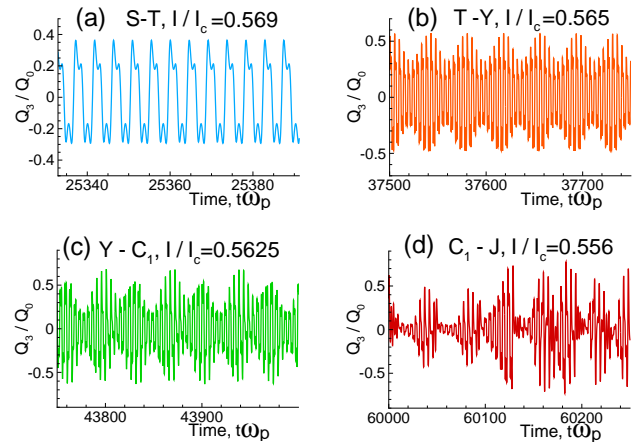}
\caption{(Color online) Charge oscillations in the third S-layer in the stack with 9 IJJ in different parts of
the BPR, specified by the value of $\emph{$I/I_c$}$ }
 \label{5}
\end{figure}
Fast Fourier transformation (FFT) analysis for the time dependence of the charge in the third S-layer presented
in Fig.~\ref{4}. The time domain samples for FFT are collected across the interval $(0, T_m)$ at each step of
the bias current. Fig.~\ref{4}a shows that before the BP at $I/I_c=0.573$ the Josephson frequency
$\omega_J=0.4542*2\pi\omega_p=2.8538\omega_p$ is observed only. In accordance with the parametric resonance,  we
can see the peak at $\omega=0.2246*2\pi\omega_p=1.4112\omega_p$ corresponding to the LPW frequency
$\omega_{LPW}$ in the $\emph{B-S}$ region (Fig.~\ref{4}b). We observe also the peak at
$\omega=0.6738*2\pi\omega_p=4.2336$ corresponding to sum of the Josephson and LPW frequencies
$\omega_J+\omega_{LPW}$. A sharp increase in the oscillation amplitude at the resonance point is observed on
other layers in the $\emph{B-S}$ region also. The oscillation in the $\emph{S-T}$ part is similar to those for
all layers in the stack excluding the layer 4. Because we consider the periodic BC along c-axis here, the layer
number is used just for distinctness. As we can see in Fig.~\ref{2}(c), the amplitude of the charge oscillation
in the layer 4 in the $\emph{B-Y}$ region is essentially smaller than that in the other layers. The FFT analysis
of the $\emph{S-T}$ part  shows the same peaks as for $\emph{B-S}$ part, but here we can see an additional  peak
$0.4395*2\pi\omega_p= 2.7615\omega_p$, which value approximately equal to 2$\omega_{LPW}$. As it is demonstrated
in Fig.~\ref{5}(b), the character of the charge oscillations in $\emph{T-Y}$ part of the BPR is different from
that in the $\emph{S-T}$ part of the BPR. The charge oscillations are characterized here by the beating process.
A specific feature of these part is the absence of the charge oscillations in the 4-th layer. At point
$\emph{Y}$ the charge oscillations in the layer 4 grow rapidly and are getting the same order as the value in
the other layers. Due to the coupling between junctions the oscillations in the other layers are increasing too,
for example, like in the third layer shown in Fig.~\ref{2}(b). The $\emph{$Y-C_1$}$ part of the BPR
(Fig.~\ref{5}(c)) is  characterized by the beating process as well, but with an  additional frequency in compare
with the $\emph{T-Y}$ region. The $\emph{$C_1-J$}$ part of the BPR presented in Fig.~\ref{5}d display a behavior
close to the chaotic one. So, the different character of the charge oscillation in the S-layers is in
correlation with the features of the CVC and reflects its nonlinear behavior in the BPR.

The question concerning the origin of the BPR and the fine structure in the CVC might be simplified if we take
into account the results obtained for stacks with even number of IJJ in Ref.\cite{prb}. In that case the LPW
with wavelength $\lambda=2d$ ($k=\pi$) is created. This wavelength of the LPW is commensurable with the length
of the stack in the c-direction and the BPR in the CVC does not appear. For such stacks with even number of
junctions the resonance region is observed only, which corresponds to the continuous increase of the amplitude
of the charge oscillations and followed by jump to the another branch.

Let us now discuss shortly the experimental situation. As we mentioned in Ref.\cite{prb}, the BPR exists in the
numerical simulations of the other authors as well. The points like breakpoint are presented in the experimental
CVC of the Refs.\cite{okanoue, suzuki} Recently, new detailed experiment was made and  the clear manifestation
of the BPR structure was found.\cite{irie} It's making promising the possibility to observe the experimentally
the features presented in this paper. The following conditions are important for that. The measurement should be
done with high precision in bias current and slow sweeping of the CVC.  Because the BPR region width is
decreasing rapidly with the number of junctions\cite{prb}, it should not be too big to observe the BPR
structure.  If the step in the bias current is big enough, it makes possible to pass the bifurcation
points,\cite{machida99} related to the parametric resonance in the system.  The influence of the inductive
coupling on the BP and BPR is not investigated yet, but if the stack's size is smaller than the Josephson
penetration depth, we may neglect it.

In summary, we predict a fine structure in the outermost branch of the current-voltage characteristics  in the
layered superconductors.  We establish a correspondence between the features of this  structure and the
character of the charge oscillations in the different superconducting layers in the stack. We demonstrate that
the different parts of the BPR correspond to the different dynamics of the charge oscillation in the
superconducting layers. Using an idea about parametric resonance at the breakpoint, we  explain the origin of
the BPR structure.   A plasma mode is a stable solution of the system in the BPR and this fact might be used in
some applications, particularly, in high frequency devices such as THz oscillators and mixers. We consider as
well that the breakpoint phenomenon might be used to understand the mechanism of the powerful coherent radiation
from the stack of the IJJ in BSCCO. We thank T. Hikihara, I. Kakeya,  A. Irie, H. J. Lee, K. Kadowaki, N. F.
Pedersen, P. Seidel, T. Koyama, M. Machida, T. Hatano, N. Plakida  for helpful discussions. Yu.M.S. thanks Prof.
J. Ishikawa and Prof. S. Noda for kind hospitality in during his stay at Kyoto university. This research was
supported by Russian Foundation for Basic Research, grant 08-02-00520-a.

\end{document}